\documentclass{ifacconf}

\usepackage{graphicx}       
\usepackage{natbib}         
\usepackage{amsmath}        
\usepackage{algorithm, algpseudocode}    
\usepackage{url}
\usepackage[final]{changes}  

\definechangesauthor[name=Ibrahim, color=blue]{ib}

\begin{document}
\begin{frontmatter}

\title{Comparison of Model Predictive and Reinforcement Learning Methods for Fault Tolerant Control}

\author[First]{Ibrahim Ahmed} 
\author[Second]{Hamed Khorasgani} 
\author[Third]{Gautam Biswas}

\address[First]{Vanderbilt University, USA (e-mail: ibrahim.ahmed@vanderbilt.edu).}
\address[Second]{Vanderbilt University, USA (e-mail: hamed.khorasgani@vanderbilt.edu).}
\address[Third]{Institute of Software Integrated Systems, Vanderbilt University, USA (e-mail:gautam.biswas@vanderbilt.edu).}

\begin{abstract}                
\replaced{A desirable property in fault-tolerant controllers is adaptability to system changes as they evolve during systems operations. An adaptive controller does not require optimal control policies to be enumerated for possible faults. Instead it can approximate one in real-time. We present two adaptive fault-tolerant control schemes for a discrete time system based on hierarchical reinforcement learning. We compare their performance against a model predictive controller in presence of sensor noise and persistent faults. The controllers are tested on a fuel tank model of a C-130 plane. Our experiments demonstrate that reinforcement learning-based controllers perform more robustly than model predictive controllers under faults, partially observable system models, and varying sensor noise levels.}{Reinforcement learning methods can derive optimal behaviour by exploration of a search space without explicit knowledge of the path to the goal states. Using functional approximations, behaviour can be generalized across partially observable state spaces. We implement reinforcement learning-based controllers and compare their performance under faults against model predictive control. Subject to state observability constraints, reinforcement learning methods demonstrate consistent and faster recovery from faults.}
\end{abstract}


\end{frontmatter}


\section{Introduction}
Complex systems are expected to operate efficiently in diverse environments. A controller that can \replaced{function}{operate} within bounds despite malfunctions and degradation in the system, or changes in the environment is safer. With the rise in automation and increasing complexity of systems, direct or timely human supervision may not always be possible. Fault Tolerant Control (FTC) seeks to guarantee stability and performance in nominal and anomalous conditions.

\replaced{\cite{bla97} delineates the difference between fail-safe and fault-tolerant control in their treatment of faults.}{Fault-tolerant control differs from fail-safe control (\cite{}) in that it provides abilities to handle performance degradation and recover from faults.} Fail-safe control employs robust control approaches to avoid degradation from faults in the first place. But it comes at the cost of additional redundancy. Fault-tolerance combines redundancy management with controller reconfiguration and adaptivity to avoid system-wide failures, while keeping the system operational (to the extent possible). \added{A fault-tolerant controller allows for graceful degradation, i.e., it may be sub-optimal after encountering faults, but it will keep the system operational for extended periods of time.}

\cite{pat97} surveys constituent research areas of FTC. We explore supervision methods in this paper. Supervision uses \emph{a priori} knowledge of system faults to choose optimal actions. Fault-tolerant supervision has been explored in several fields including probabilistic reasoning, fuzzy logic, and genetic algorithms. This paper focuses on the evaluation of two supervision approaches for FTC: model predictive (MP) and reinforcement learning (RL)-based control. We assume system models are reasonably accurate, the fault diagnoser is correct, and the system can reconfigure in response to the supervisor.

MP and RL-based control rely on sampling \replaced{the state space in real-time to find a control policy.}{future states using a dynamic system model to select an optimal action.} However, such FTC approaches may be subject to additional constraints due to faults \added{and constraints on operation}. Controllers can be limited by processing power, poor observability of states \replaced{due to}{because of} sensor failures, and loss of functionality because of faulty components. We explore how the \replaced{two control approaches fare under various conditions.}{different approaches fare under various sampling restrictions.}

\added{Some work has been done in designing RL-based fault adaptive controllers. \cite{lewis2012reinforcement} provides a theoretical discussion of using RL in real-time to find a control policy that converges to the optimum. \cite{liu2017adaptive} applies RL based control to a discrete time system and uses a fully connected shallow neural network to approximate the control policy.}

\added{The main contribution of this work is to benchmark RL-based controllers against MP control.} \replaced{We}{In this paper, we} implement discrete-time MP control and propose two approaches for applying RL principles towards FTC. The controllers are tested on a model of fuel transfer system of a C-130 cargo plane where the controllers attempt to maintain \added{a}  balanced fuel distribution in the presence of leaks.

The rest of the paper is organized as follows. Section 2 provides a background of MP and RL-based control. Section 3 discusses online and offline RL controller design. Our case study is presented in Section \ref{casestudy}. The results are presented and discussed in Section \ref{results}. Sections \ref{future} and  \ref{conclusions} \replaced{discuss future work, open problems in this domain, and the conclusions.}{presents the conclusion.} 

\section{Background}

\subsection{Model Predictive Control}

Model Predictive Control (MPC) \deleted{is a process control method that} optimizes the current choice of action \added{by a controller} that drives the system state towards the desired state. It does this by \replaced{modeling}{predicting} future states of 
the system+environment and a finite receding horizon over which to predict state trajectories and select the ``best'' actions \deleted{at every time step}. At each time step, \replaced{a sequence of actions is selected up until the horizon}{actions are selected in the horizon} such that a cost or distance measure to a desired goal is minimized. For example, for a system with state variables $(x_1,x_2,...,x_N)$ and the desired state $(d_1,d_2,...,d_N)$ the cost may be a simple Euclidean distance measure between the two vectors.

A model predictive controller \replaced{operates over a}{has an available} set of states $s \in S$ \replaced{,}{and} actions $a \in A$, and a \added{state transition} model of the system $T:S \times A \rightarrow S$. It generates a tree of state trajectories rooted at the system's current state. The tree has a depth equal to the \deleted{pre-}specified  lookahead horizon for the controller. The shortest path trajectory, \replaced{that is}{i.e.}, the sequence of states and actions producing the smallest cost is chosen, and the first action in that trajectory is executed. The process repeats for each time step.

\cite{gar89} discuss MPC in further detail particularly in reference to continuous systems. \cite{abd05} presents a case study for MPC of a hybrid system. The control algorithm for the discrete subsystem is \added{essentially} a limited breadth-first search of state space. \replaced{They}{Abdelwahed et al.} use a distance map, \deleted{as a cost function} which \replaced{uses the euclidean distance between the controller state and the closest goal state as the cost function.}{returns the distance from the closest goal state.} The \replaced{MPC}{control} algorithm implemented for this work \replaced{is adapted from \cite{abd05} and described in Algorithm \ref{alg:MPC}.}{ searches all reachable states at every time step to pick the optimal action. The MPC algorithm is summarized as Algorithm \ref{alg:MPC} below.}

\begin{algorithm}[ht]
\caption{Model Predictive Control}
\label{alg:MPC}
\begin{algorithmic}[1]
\Require system model $T:S \times A \rightarrow S$
\Require distance map $D:S \rightarrow \Re$
\Require horizon $N$
\State $s_{0} \gets CurrentState$
\State Initialize state queue $Queue=\{s_0\}$
\State Initialize state set $Visited=\{s_0\}$
\State Initialize optimal state $Optimal \gets Null$
\State $MinDistance \gets \infty$
\While{$Queue.length>0$}
    \State $state=Queue.pop()$
    \If{$state.depth > N$}
        \State Break
    \EndIf
    \State $newStates = state.neighbourStates \cap \neg Visited$
    \For {$s \in newStates$}
        \State Visited.add(s)
        \State Queue.insert(s)
        \If{$D(s)<MinDistance$}
            \State $MinDistance=D(s)$
            \State $Optimal=s$
        \EndIf
    \EndFor
\EndWhile
\While{$OptimalState.previous \neq s_0$}
    \State $Optimal \gets Optimal.previous$
\EndWhile
\State $action=T(Optimal.previous, \cdot) \rightarrow Optimal$
\end{algorithmic}
\end{algorithm}

\subsection{Reinforcement Learning}


Reinforcement learning is the learning of behaviour by an agent, or a controller, \added{from feedback} through repeated interactions with its environment. \cite{kae96} divide RL into two broad approaches. In \replaced{the first, the genetic programming approach}{the first approach}, an agent explores different behaviours to find ones that yield better results. \replaced{In the second, the dynamic programming approach,}{In the second approach,} an agent estimates the value of taking individual actions from different states in the environment. The value of each state and action dictates how an agent behaves. This section explores methods belonging to the latter approach.

An agent (controller) is connected to (interacts with) the system+environment through its actions and perceptions. Actions an agent takes cause state transitions of the system in the environment. The change is perceived as a reinforcement signal from the system+environment, and the agent’s measurement of the new state.

The standard RL \replaced{problem}{model} can be represented as a Markov Decision Process (MDP). A MDP consists of a set of states $S$, a set of actions $A$, a reward function \replaced{$R:S \times A \times S \rightarrow \Re$}{$R:S \times S \rightarrow \Re$,} which provides reinforcement after each state change, and a state transition function $T:S \times A \rightarrow \Pi(S)$, which determines the probabilities of going to each state after an action. This is the model for the system operating in an environment. For the purposes of this paper, the transition function is assumed to deterministically map $s \in S$ and $a \in A$ to the next state $s' \in S$.

In a MDP, everything an agent needs from the environment is encoded in the state. The history of prior states does not affect the value or reward of the following states and actions. This is known as the Markov property.

An agent's objective is to maximize its total future reward. \replaced{The value of a state is the maximum reward an agent can expect from that state in the future.}{The maximum expected rewards from a state are its value:}

\begin{equation} \label{eq:value}
V(s) = \max_{a \in A} ( E[R(s,a,s') + \gamma V(s')] ),
\end{equation}

where $\gamma$ is the discount factor that \replaced{weighs delayed rewards against immediate reinforcement.}{determines how future rewards are weighed against immediate reward.}
Alternatively, an agent can learn the value of each action, the q-value, from a state:

\begin{equation} \label{eq:qvalue}
Q(s,a)=E[R(s,a,s') + \gamma \max_{a \in A} Q(s', a)].
\end{equation}

 We use the q-value formulation for this paper. An agent can exploit either value function to derive a policy that governs its behaviour during operation. To exploit the learned values, a policy greedily or stochastically selects actions that lead to states with the highest value:

\begin{equation} \label{eq:policy}
\begin{aligned}
\pi_{greedy}(s) & = arg \max_{a \in A} Q(s,a)\\
\pi_{stochastic}(s) & = \Pi(A)
\end{aligned}
\end{equation}

\replaced{\eqref{eq:value}, known as the Bellman equation, and \eqref{eq:qvalue} can be solved recursively for each state using dynamic programming. However, the complexity of the solution scales exponentially with the number of states and actions. Various approaches have been proposed to approximate the value function accurately enough with minimal expenditure of resources.}{Various methods for calculating value and policy functions are surveyed in \cite{}. The recursive form of the value function can be solved through dynamic programming (DP). This requires traversal of the entire state space accessible from $s$. However, for large state spaces this approach becomes intractable.}

Monte Carlo (MC) methods (\cite{sut17}) alleviate the Curse of Dimensionality by approximating the value function. \replaced{Instead of traversing the entirety of state-space, they}{They} generate episodes of states connected by actions. For each state in an episode, the total discounted reward received after the first occurrence of that state is stored. The value of the state is then the average total reward across episodes. Generating an episode requires a choice of action for each state. The exploratory action selection policy \replaced{$\pi_{exp}$}{$\pi$} can be uniform, partially greedy with respect to the highest valued state ($\epsilon$-greedy), or proportional to the relative values of actions from a state (softmax). The MC approach still requires conclusion of an episode before updating value estimates. Additionally, it may not lend itself to problems that cannot be represented as episodes with terminal or goal states.

Temporal Difference (TD) methods provide a compromise between DP and MC approaches. Like MC, they learn episodically from experience and use different action selection policies during learning to \replaced{explore the state space}{facilitate state space exploration}. Like DP they do not wait for an episode to finish to update value estimates. \deleted{In its simplest form, }TD updates the value function iteratively \replaced{at each time step $t$.}{ after a single time step:}

\begin{equation}
\begin{aligned}
G_{t} &=  R_{t+1} + \gamma \max_{a \in A} Q_{t-1}(s_{t+1},a) \\
Q_{t}(s_{t}, a_{t}) &= Q_{t-1}(s_{t}, a_{t}) + \alpha (G_{t} - Q_{t-1} (s_{t},a_{t})),
\end{aligned}
\end{equation}

where $\alpha$ is the learning rate for the value function. \deleted{The subscripts represent function or variable evaluation at time, $t$.} \replaced{$G_{t}$,}{Here, $G_{t}=R_{t+1} + \gamma \max_{a \in A} Q_{t-1}(s_{t+1},a)$,} called the return, is \replaced{a}{the} new estimate for the value. \replaced{The value function is updated using the error between the new estimate and the existing approximation.}{$G_{t}-Q_{t-1} (s_{t},a_{t})$ is the error between the current and previous estimates of $Q(s,a)$.} Note that $G$ estimates the new value by only using the immediate reward and backing up the discounted previous value of the next state. This is known as $TD(0)$. TD methods can be expanded to calculate better estimates of values by explicitly calculating rewards several steps ahead, and only backing up value estimates after that. These are known as $TD(\lambda)$ methods. For example, the return for \replaced{$TD(2)$}{TD(2)} is:

\begin{equation}
\begin{aligned}
G_{t}^{(2)} &= R_{t+1} + \gamma (R_{t+2} + \gamma \max_{a \in A} Q_{t-1}(s_{t+2}, a)) \\
            &= R_{t+1} + \gamma R_{t+2} + \gamma^{2} \max_{a \in A} Q_{t-1}(s_{t+2}, a) \\
Q_{t}(s_{t}, a_{t}) &= Q_{t-1}(s_{t}, a_{t}) + \alpha (G_{t}^{(2)} - Q_{t-1}(s_{t}, a_{t}))
\end{aligned}
\end{equation}

In the extreme case, when $G$ is calculated for all future steps until a terminal state \replaced{$s_T$}{$s_{Terminal}$}, the TD method becomes a MC method because value estimates then depend on the returns from an entire episode.

\replaced{An}{A further} improvement on $TD(\lambda)$ methods is achieved by \replaced{making the return $G$ more representative of the state space. Values of actions are weighed by the agent's action selection policy.}{weighing value backups by the probability of taking the corresponding action at each time step during lookahead.} This is known as $n-$step Tree Backup (\cite{pre00}). \deleted{For each step in the lookahead, the return is the sum of the weighed value backups of available actions not taken, and the discounted return of the action taken.} Defining \replaced{$\pi_{exp}(a | s)$}{$\pi(a | s)$} as the probability of \replaced{exploratory actions}{taking action $a \in A$ from $s \in S$}, \replaced{$s_T$}{$s_{Terminal}$} as the terminal time, and:

\begin{equation}
\begin{aligned}
V_{t} &= \sum_{a} \pi_{exp}(a | s_{t}) Q_{t-1}(s_{t}, a) \\
\delta_{t} &= R_{t+1} + \gamma V_{t+1} - Q_{t-1}(s_{t}, a_{t}),
\end{aligned}
\end{equation}

where $V_t$ is the expected value of $s_t$ under the action selection policy \replaced{$\pi_{exp}$}{$\pi$}, and $\delta_t$ is the error between the return and prior \deleted{q$-$}value estimate. Then the $n^{th}$-step return is given by:

\begin{equation}
\begin{aligned}
G_{t}^{(n)} &= Q_{t-1}(s_{t}, a_{t}) + \\
&\sum_{k=t}^{\min(t+n-1, T-1)}\delta_{k} \prod_{i=t+1}^{k}\gamma \pi_{exp}(a_{i} | s_{i}).
\end{aligned}
\end{equation}

The \replaced{value function}{q$-$value} can then be updated using the error $G_{t}^{(n)}-Q_{t-1}(s_t,a_t)$. A longer lookahead gives a more representative estimate of value. A greedy action selection policy with $n=0$ reduces this to the $TD(0)$\deleted{ case discussed previously}.

The following section presents a modified $n-$step Tree Backup algorithm tailored for adaptive fault-tolerant control.
\section{Reinforcement Learning-based controller design}

Temporal difference based RL methods \added{iteratively} derive a controller's behaviour by estimating the value of states \replaced{and actions}{(and actions)} in a system through exploration. During operation, the controller exploits the knowledge gained through exploration by selecting actions \replaced{with the highest value}{that yield the highest reward values}. This works well if the system dynamics is stationary. The agent is able to \replaced{achieve optimal control}{accrue knowledge} by exploring over multiple episodes and iteratively \replaced{converging}{converge} on the value function. Conservative learning rates can be used such that the value approximation is representative of the agent's history.

When a fault occurs in a system, its dynamics change. \replaced{The extant value function}{Calculated values of states} may not reflect the most rewarding actions in the new environment model. The agent seeks to optimize actions under the new system dynamics. The agent must, therefore, estimate a new value function by exploration every time a fault occurs. The learning process can be constrained by deadlines and \replaced{sensor noise}{observability of the environment}. This paper proposes design of a RL-based controller subject to the following requirements:

\begin{description}

\item \emph{Adaptivity}: The \replaced{controller}{agent} should be able to \replaced{remain functional under previously unseen faults}{handle faults beyond a single FTC scheme} (\cite{gol93}). \deleted{RL methods have an advantage over MP control here. The value function is estimated at run time, whereas the cost metric for MP control is a design time parameter which may not be accommodating of faults.}

\item \emph{Speedy convergence}: The agent’s behaviour should be responsive to faults as they occur.

\item \emph{Sparse sampling}: The system model may lose reliability following a fault \added{due to sensor noise or incorrect diagnosis}. \deleted{Or the agent may be under time constraints.} The agent should be able to calculate value estimates representative of its \added{local} state space \deleted{neighbourhood} from minimal (but sufficient) sampling. 

\item \emph{Generalization}: The agent should be able to generalize its behaviour over states not sampled in the model during learning.
	
\end{description}
	
The temporal difference RL approach is inherently adaptive \deleted{to environmental changes} as it \replaced{frequently updates its policy from exploration.}{relies on results of exploration.} Its responsivity can be enhanced by increasing the learning rate $\alpha$. In the extreme case, $\alpha=1$ replaces the last value of \replaced{an}{a state and} action with the new estimate at that time step. However, larger values of $\alpha$ may not converge the value estimate to the \replaced{global optimum.}{globally optimal values.}

RL controllers have to balance \replaced{exploratory actions to discover new optima in the value function against exploitative actions that use the action values learned so far.}{exploration of the state space against exploitation of learned values at each step during operation.} An exploration parameter $\epsilon \in [0,1)$ can be set, which determines the periodicity of value updates and hence the controller's responsivity \added{to system faults}.

\deleted{RL approaches are compatible with sparse state sampling. TD learning methods do not need to experience an entire episode until a terminal state to get value estimates. Like MP control, they can have a finite receding horizon. Unlike MP control, they can update value estimates of all traversed states. This means that the learning process can halt once sufficient experience has accumulated. In the best-case scenario, an RL controller re-learns a value function only after fault occurrences.}

Hierarchical RL can be used to sample the state space at various resolutions (\cite{lam10}). An agent can use a hierarchy function $H:S \rightarrow \Re^{+}$ that gives the step size \replaced{of actions an agent takes}{a system model uses} to sample states. $H$ can yield smaller time steps for states closer to goal and coarser sampling for distant states. This allows the agent to \replaced{update the value function more frequently with states where finer control is required.}{look farther under the same state sample size.}

TD RL methods can use \textit{function approximation} to generalize the value function over the state space instead of using tabular methods. Gradient descent is used with the error computed by the RL algorithm to update parameters of the value function provided at design time. Function bases like radial, Fourier, and polynomial can be used to generalize a variety of value functions.

Algorithm~\ref{al:nstep} implements Variable $n-$step Tree Backup for a single episode. It updates the value estimate after exploring a sequence of states up to a finite depth $d$ \added{of actions} with varying \replaced{step sizes}{resolution}. Value estimates for each state in the sequence depend on following states up to $n$ steps ahead. A RL-based controller explores multiple episodes to converge on a locally optimum value for actions. During operation it exploits the value function to pick the most \replaced{valuable}{desirable} actions as shown in \eqref{eq:policy}.

The following subsections discuss two approaches to approximating values for a RL-based controller.

\begin{algorithm}[ht]
\caption{Variable n-Step Tree Backup Episode} \label{al:nstep}
\begin{algorithmic}[1]
\Require initial state $s_0$
\Require system model $T:S \times A \rightarrow S$
\Require value function $Q:S \times A \rightarrow \Re$
\Require reward function $R:S \times S \rightarrow R$
\Require hierarchy $H:S \rightarrow \Re$
\Require action selection policy $\pi:S \rightarrow \Pi(A)$
\Require parameters: $\alpha, d, \gamma, n$
\State Get action $a_0$ from $\pi_{exp}(s_0)$
\State $Q_{0} \gets Q(s_0,a_0)$
\State $T \gets \infty,t \gets 0, \tau \gets 0$
\While{ $t < d$ and $\tau < T-1$}
    \If{$t < T$}:
        \State Take action $a_t$ for $H(s_t)$ steps
        \State $s_{t+1} \gets T(s_{t},a_{t})$
        \State $R_{t} \gets R(s_{t},a_{t},s_{t+1})$
        \If{$s_{t+1}$ is in goal states}
            \State $T \gets t+1$
            \State $\delta_{t} \gets R_{t}-Q_{t}$
        \Else:
            \State $\delta_{t} \gets R_t+\gamma \sum_{a} \pi_{exp}(a \mid s_{t+1})Q(s_{t+1},a)-Q_{t}$
            \State $a_{t+1} \gets \pi_{exp}(s_{t+1})$
            \State $Q_{t+1} \gets Q(s_{t+1},a_{t+1})$
            \State $\pi_{t+1} \gets \pi_{exp}(a_{t+1}\mid s_{t+1})$
        \EndIf
    \EndIf
    \State $\tau \gets t-n+1$
    \If{$\tau \geq 0$}
        \State $E \gets 1$
        \State $G \gets Q_t$
        \For{$k=\tau,...,\min(\tau + n - 1, T - 1)$}
            \State $G \gets G + E \cdot \delta_{k}$
            \State $E \gets \gamma E \cdot \pi_{k+1}$
        \EndFor
        \State $Error \gets Q(s_\tau,a_\tau) - G$
        \State Update $Q(s_\tau,a_\tau)$ with $\alpha \cdot Error$
    \EndIf
    \State $t \gets t + 1$
\EndWhile
\end{algorithmic}
\end{algorithm}

\subsection{Offline Control}

An offline RL-based controller derives the value function by dense and deep sampling of the state space. The learning \replaced{is done once each time a new value approximation is required.}{process takes place only once before operation.} \replaced{The agent learns from episodes over a larger}{Episodes are called over a higher resolution} sample of states \added{in the model}. Each episode \replaced{lasts till a terminal state is reached.}{is allowed to continue till termination.} \deleted{The controller exploits the learned value function for the duration of its operation until a new fault occurs and system dynamics change.} Offline control is computationally expensive. However, by employing a partially greedy exploratory action selection policy $\pi$ and a suitable choice of learning rate $\alpha$, it is liable to converge to the global optimum.

\begin{algorithm}[ht]
\caption{Offline RL Control}
\begin{algorithmic}[1]
\State $d \gets \infty$
\For{$s \in S$}
    \State Call Episode
\EndFor
\While{True}
    \State $s_{0} \gets CurrentState$
    \State $action = arg \max_{a \in A} Q(s_{0},a)$
\EndWhile
\end{algorithmic}
\end{algorithm}

\subsection{Online Control}

Online RL-based control interleaves operation with multiple shorter learning phases. The agent sporadically explores via limited-depth episodes starting from \replaced{the controller's current state}{states in the agent's state-space neighbourhood}. The sample size is small and local. Each exploratory phase is comparatively inexpensive but may only converge to a local optimum.

\begin{algorithm}[ht]
\caption{Online RL Control}
\begin{algorithmic}[1]
\Require exploration rate $\epsilon$
\Require sampling density $\rho$
\While{True}
    \State $s_{0} \gets CurrentState$
    \If{$RandomNumber < \epsilon$}
        \State $neighbours \gets \{T(s_{0},a):a \in A\}$
        \For{$state \in neighbours$}
            \If{$RandomNumber < \rho$}
                \State Call Episode
            \EndIf
        \EndFor
    \EndIf
    \State $action = arg \max_{a \in A} Q(s_{0},a)$
\EndWhile
\end{algorithmic}
\end{algorithm}

\replaced{Like}{In other words, like} MPC, online RL employs a limited lookahead horizon \replaced{periodically to learn the best actions. Unlike MPC, online RL remembers and iteratively builds upon previously learned values of actions.}{, but the difference is that it continues to learn and update its value functions through each iterative limited lookahead phase.}
\section{Case Study}
\label{casestudy}
\begin{figure*}[ht]
\begin{center}
\includegraphics[width=16.4cm]{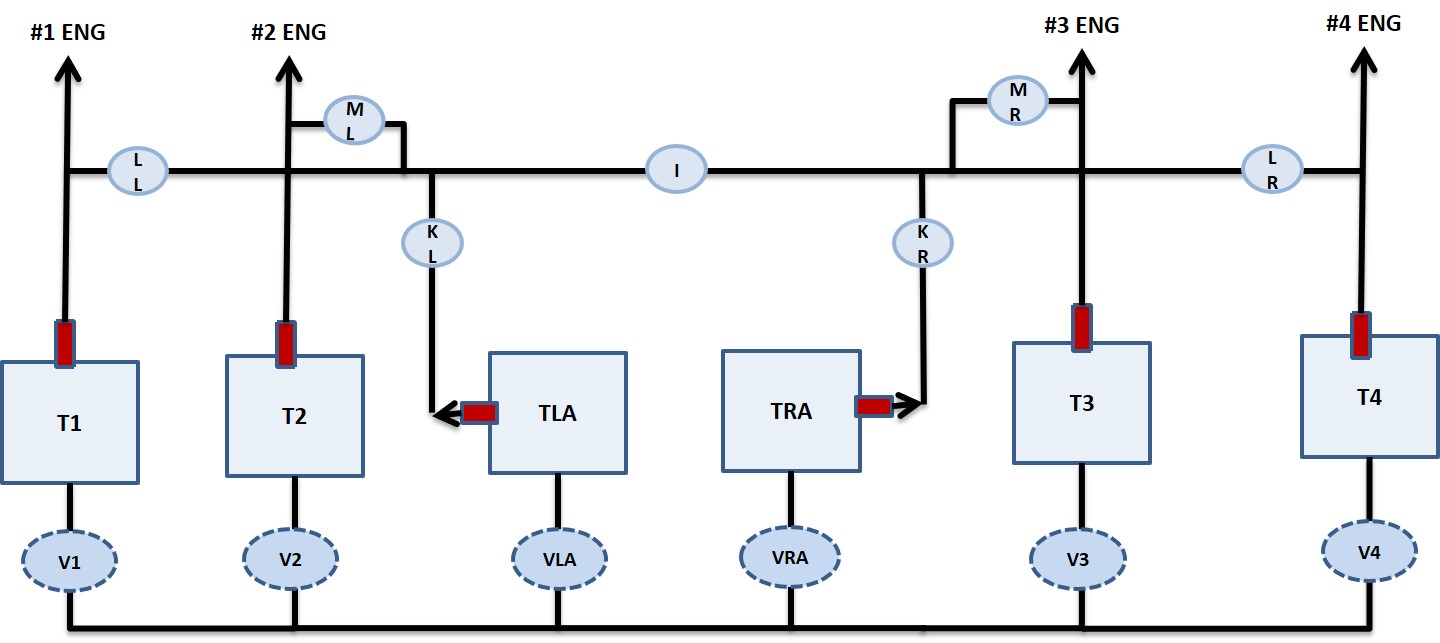}    
\caption{Simplified C-130 fuel system schematics. \replaced{The controller manages valves (dotted ovals)}{Controllers manage valve states} and can observe fuel tank levels. \replaced{Net outflow}{Outflow} to engines via pumps \added{(solid ovals)} is \replaced{controlled independently.}{pre-determined.}}
\label{fig:fuelsystem}
\end{center}
\end{figure*}

The three control schemes (MP, Offline RL, and Online RL) are applied to a simplified version of the fuel tank system of a C-130 cargo plane (see  Fig.~\ref{fig:fuelsystem}). There are six fuel tanks. The fuel tank geometry is symmetric and split into a left/right arrangement\deleted{,} where  each side primarily feeds one of two engines. Control of fuel transfer is required to maintain aircraft center of gravity (CG). Changes in the fuel pump operating performance, plumbing, engine demand, and valve timing can affect the way each tank transfers \added{fuel}. The CG is a function of fuel quantities in the tanks, therefore,  the fuel control system can restore the CG by turning transfer valves off or on to bring the system back into acceptable limits.

Fuel transfer is controlled by two components: pumps and valves. Pumps maintain a constant flow to the engines. Under nominal operating conditions, the fuel transfer controller is designed to pump fuel from outer tanks first, and then sequentially switch to inner tanks. Each tank has a transfer valve feeding into a shared conduit. When valves are opened, pressure differentials between tanks may result in fuel transfer to maintain the balance.

The state of the system is described by twelve variables: six fuel tank levels, which are continuous variables, and six valve states, that are discrete-valued. For simplicity, fuel levels are represented on a $0-100$ min-max scale. Valve positions are binary on/off variables. A controller can switch any combination of valves on or off, giving a total of 64 possible configurations fr the action space.

\begin{equation}
\begin{aligned}
s &: \{T_{1}, T_{2}, T_{LA}, T_{RA}, T_{3}, T_{4}, V_{1}, V_{2}, V_{LA}, V_{RA}, V_{3}, V_{4}\}\\
a &: \{V_{1}, V_{2}, V_{LA}, V_{RA}, V_{3}, V_{4}\}
\end{aligned}
\end{equation}

Faults in the fuel tank system are leaks in fuel tanks. When a leak occurs, a tank drains additional fuel at a rate proportional to the quantity left. It is assumed that controllers have accurate system models and the fault diagnosis results are provided in a timely manner. WE make the single fault assumption, i.e., at any point in time, there is at most one fault in the system.

\added{Sensor noise is simulated by scaling the measurements of tank levels with values derived from a Gaussian distribution centered at $1$ with a standard deviation $\sigma$. The noise is assumed to be inherent to the sensors.}

A trial begins with all tanks filled to capacity with a possible fault in one tank. The trial concludes when no fuel is left in tanks. For each trial, the maximum imbalance and the time integral of imbalance are used as performance metrics.

Goal states in the system are configurations where there is zero moment about the central axis. The magnitude of the moment is the imbalance in the system. The distance map $D$ used by MP controller is a measure of imbalance.

\begin{equation}
    D(s) = Abs(3 (T_{1} - T_{4}) + 2 (T_{2} - T_{3}) + (T_{LA} - T_{RA}))
\end{equation}

RL controllers are supplied with a reward function $R$. The controller is rewarded for reaching low-imbalance states and for doing it fast when there is more fuel to spare.

\begin{equation}
\begin{aligned}
R(s,s') &= \frac{T'_{1} + T'_{2} + T'_{LA} + T'_{RA} + T'_{3} + T'_{4}}{600} \\
        &+ \frac{1}{1 + D(s')}
\end{aligned}
\end{equation}

RL controllers also hierarchically sample state space. The further a state is from goal, the more imbalanced the tanks are, therefore, the longer the action duration is at that state.

\begin{equation}
H(s) = 1 + Log_{10}(1 + D(s))
\end{equation}

Value function approximation is done via a linear combination $Q(s,a)$ of normalized fuel level and valve state products with a bias term. At each episode, the $Error$ is used to update weights $\vec{w}$ by way of gradient descent.

\begin{equation}
\begin{aligned}
Q(s, a) &= \vec{w} \cdot \Big\{ \frac{T_{1}(1+V_{1})}{200},\ldots, \frac{T_{4}(1+V_{4})}{200}, 1 \Big\}^T \\
\nabla Q_{w} &= \Big\{ \frac{T_{1}(1+V_{1})}{200},\ldots, \frac{T_{4}(1+V_{4})}{200}, 1 \Big\}^T \\
\vec{w} &\gets \vec{w} - \alpha \cdot Error \cdot \nabla Q_{w}
\end{aligned}
\end{equation}

RL controllers employ a softmax action selection policy \replaced{$\pi_{exp}$}{$\pi$} during exploration~(\cite{doy02}). \replaced{The probability of actions}{Their choice of exploratory actions} is proportional to the values of those actions at that instant. This ensures that during learning, states which yield more reward are prioritized for traversal.
\section{Results \added{and discussion}}
\label{results}

For each of the three control schemes, \deleted{and for a case where no control was implemented,} thirty trials were carried out with random faults (\cite{ahm17}). \deleted{Gaussian noise ($\sigma=0.5$) was added to each observation of fuel tank levels to simulate model unreliability due to faults.} Table~\ref{tb:params} shows the values of the baseline parameters used. For these parameters, MP and online RL controllers sampled the same number of states at each step on average.

\begin{table}[ht]
\begin{center}
\begin{tabular}{lcccc}
        Name                & Symbol        & MP    & Offline RL    & Online RL  \\    \hline
        Learning rate       & $\alpha$      & N/A   & 0.1           & 0.1 \\
        Discount            & $\gamma$      & N/A   & 0.75          & 0.75 \\
        Depth               & $d$           & N/A   & 30            & 5 \\
        Lookahead steps     & $n$           & N/A   & 10            & 5 \\
        Exploration rate    & $\epsilon$    & N/A   & N/A           & 0.4 \\
        Sampling density    & $\rho$        & 1     & 1             & 0.5 \\
        Horizon             & $N$           & 1     & N/A           & N/A \\            \hline \\
\end{tabular}
\caption{Controller baseline parameters} \label{tb:params}
\end{center}
\end{table}

\begin{figure}[ht]
    \centering
    \includegraphics[width=8.2cm]{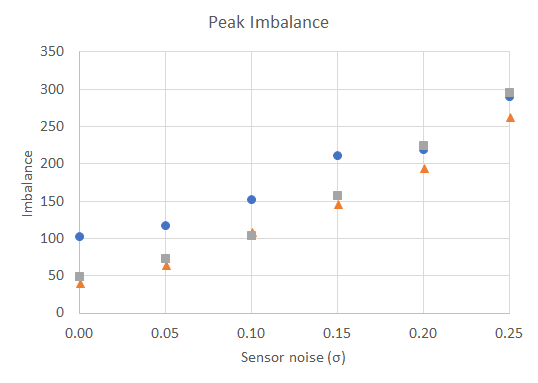}
    \includegraphics[width=8.2cm]{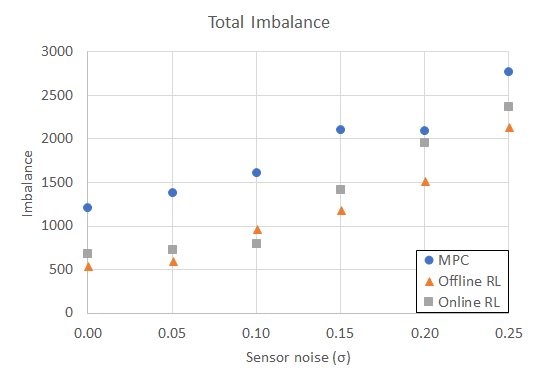}
    \caption{Performance of controllers under varying sensor noise. RL-based controllers are more robust to sensor noise than MPC. Lower imbalances are better.}
    \label{fig:noise}
\end{figure}

Under baseline conditions, \replaced{Figure~\ref{fig:noise}}{Table~\ref{tb:baseline}} shows the performance of various controllers \added{using two metrics}. \added{(1) \textit{Maximum imbalance} is the average maximum value $D(s)$ reached during trials. (2) \textit{Total imbalance} is the average time integral of $D(s)$ over each trial.} \replaced{}{All controllers exhibited an improved performance over the unsupervised case. For the trials with no noise, model predictive (MP) control produced the best recovery from faults, followed by offline, and finally online reinforcement learning \deleted{(RL)}-based control.}

These results are representative of how controllers sample the state space to select actions. MP control samples all reachable states within a horizon at each step. It finds the locally optimal action. \added{The choice of each action depends on the instantaneous values associated with the states in the model. In case of sensor noise, sub-optimal actions may be chosen.}

Offline RL control samples a \added{large} fraction of potential successor states once, \replaced{and derives a single value function and control policy that is applied to select actions in the future.}{derives a value function, and bases behaviour on it throughout.} Online RL periodically samples a \added{small, local} fraction of successor states \deleted{in the neighbourhood} to make recurring corrections to its policy. \added{It starts off being sub-optimal, but with each exploratory phase improves on its choice of actions till it converges to the lowest cost choices. It is possible in scenarios where response time is critical, online RL may not have time to sample enough states to converge to an acceptable policy.}

\added{Both RL methods rely on accumulated experience. Having a learning rate means that state values, and hence the derived control policy, depend on multiple measurements of the model. For Gaussian sensor noise, the average converges to the true measurement of the state variable. Therefore, the controller is influenced little by random variations in measurements. Furthermore, the choice of value function can regularize the policy derivation so it does not over-fit to noise.}

\deleted{Both RL methods are sensitive to the choice of the reward function. In contrast to MP control, where the distance map to the goal is a design time parameter, RL methods do not need any knowledge of goal states. Instead, they take the most rewarding path to recovery, \replaced{which}{and this} may not be the shortest path.}

RL methods are also sensitive to the form of the value approximation function, which \replaced{may under-fit the true value of states or over-fit to noise.}{may not correspond to the true value of states.} They forego some accuracy in state valuation by using approximations in exchange for \replaced{the ability to interpolate values for unsampled sates.}{better interpolability for unsampled states. These  two dependencies explain the inferior performance relative to MP control.} The use of neural networks as general function approximators may allow for a more accurate representation of true state values\deleted{(\cite{cyb89})}, and therefore, better control.

\deleted{Conversely, for the case of \replaced{sensor}{fault-induced} noise, MP control showed the worst performance whereas RL methods were consistent. For MP control, the choice of each action depends on the instantaneous measurement of states in the model. In case of random noise, sub-optimal actions will be chosen. RL methods, on the other hand, rely on accumulated experience. Having a learning rate means that state values, and hence the derived control policy, depend on multiple measurements of the model. For the case of Gaussian noise, the average \deleted{of disturbances} converges to the true measurement of the state variable. Therefore, the controller is influenced little by random variations. Furthermore, the choice of value function can \replaced{regularize the policy derivation so it does not over-fit to noise.}{have a smoothing effect to ensure that random noise does not cause abrupt changes in values between states.}}

\begin{figure}[ht]
    \centering
    \includegraphics[width=8.2cm]{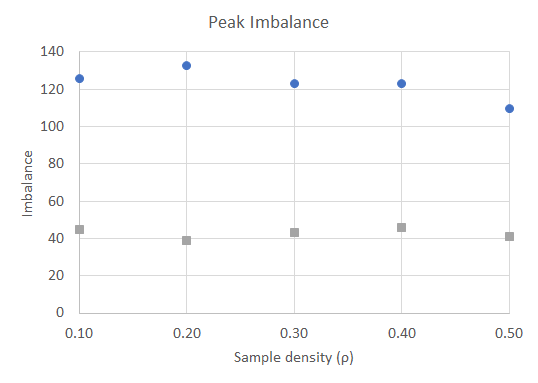}
    \includegraphics[width=8.2cm]{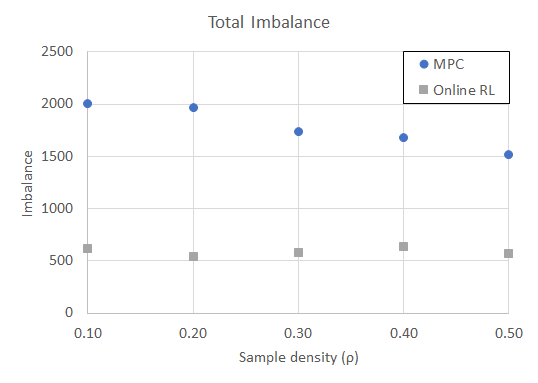}
    \caption{Performance of controllers under varying sample densities. RL-based controllers are more robust to smaller state samples than MPC. Lower imbalances are better.}
    \label{fig:density}
\end{figure}

Another set of \added{30} trials was carried out with online RL \added{and MP controllers} to simulate sampling and processing constraints \added{during operation}. State sample sizes were restricted by lowering the sampling density $\rho$. \added{At each step, the MP and RL controllers could only advance a fraction $\rho$ of available actions to their neighbouring states to explore.} The exploration rate was set to $\epsilon=0.2$ \added{to ensure equal sample sizes for both controllers. Sensor noise was reset to $\sigma=0$}. \replaced{Results are shown in Figure~\ref{fig:density}.}{Results are tabulated in Table~\ref{tb:rl-density}.}

\added{Online RL control is quick to recover from a fault and maintains smaller imbalance for the duration of the trial. By having a value function approximation, the controller estimates utilities for actions it cannot sample in the latest time step based off of prior experience.  MP control, however, is constrained to choose only among actions it can observe in the model. It is, therefore, more likely to make suboptimal choices. With a decreasing sample density, MP control's performance deteriorates whereas online RL control is less sensitive to the change.}

\deleted{For comparison, a number of trials using MP controller were carried out with decreasing sampling densities. At each step, the MP controller could only see a fraction $\rho$ of its neighbouring states to explore. On average both MP and RL controllers had the same state sample sizes. Table~\ref{tb:mp-density} documents the results.}


\deleted{Of note is the insignificant change in online RL control performance as the sample size is reduced by $80\%$. For MP control, however, the performance is worse and deteriorates faster with decreasing sampling density. Indeed performance is consistently poorer than the case where no control was implemented.}

Finally, combining \replaced{sensor noise}{random model noise} and low sampling densities yielded \replaced{a significant}{the largest} disparity in performance between the RL and MP control approaches as shown in Table~\ref{tb:combined}.

\begin{table}[ht]
\begin{center}
\begin{tabular}{lcc}
        Control            & Max Imbalance & Total Imbalance  \\   \hline
        MP                 & 147.66         & 1972.89 \\
        Online RL          & 75.17          & 721.57 \\           \hline \\
\end{tabular}
\caption{MP vs Online RL control \newline ($\sigma=0.05, \rho=0.5$)} \label{tb:combined}
\end{center}
\end{table}

The results show that reinforcement learning-based control is more robust. When the model is \replaced{insufficiently}{partially} observable, either due to \replaced{sensor}{random} noise or due to sampling constraints, RL controllers are able to generalize behaviour and deliver \replaced{consistently better performance than MP control.}{consistent performance.} The complexity of both MP and RL approaches is linear in the number of states sampled. However, by discarding prior experience and sampling states anew at each step, model predictive control forfeits valuable information about the environment \added{that RL methods exploit}.
\section{\added{Future work}}
\label{future}

\added{The fuel tank model made a simplifying assumption of discrete-time dynamics. The problem of continuous-time reinforcement learning-based control remains open and has been discussed in some detail by \cite{doya2000reinforcement} and \cite{lewis2012reinforcement}.}

\added{The RL-based controllers approximated the value of actions based on a second order polynomial function, which was learned using stochastic gradient descent. The universal approximation theorem as described by \cite{cyb89} shows that single hidden-layer neural networks with sigmoidal non-linearities can approximate any continuous and real-valued function under some constraints. Using such a basis for the value function may lead to more optimal control policies that reflect the true values of actions.}

\added{Using stochastic gradient descent is computationally efficient as a single observation is used to compute changes to the value approximation. However, it is prone to instability in convergence in case of noisy data. Experience replay, as applied to control systems by \cite{adam2012experience}, maintains a history of prior observations which can be randomly sampled in batches to calculate value updates. By aggregating observations, the effects of sampling errors and noise may be mitigated and an optimal control policy may be derived sooner.}
\section{Conclusion}
\label{conclusions}

Reinforcement learning-based control provides \replaced{a competitive}{an} alternative to model predictive control, especially in situations when the system degrades over time, and faults may occur during operations. RL control's independence of explicitly defined goals allows it to operate in environments where faults may render nominal goal states \replaced{unfeasible}{unstable or unreachable}. The ability to generalize behavior \added{over unseen states and actions, and to derive control policies from accumulated experience,} \deleted{and perform consistently under low sampling rates} make RL control \replaced{the preferred}{a good} candidate for \replaced{system models subject to computational constraints, partially observable environments, and sensor noise. There are several venues for exploration in the choice of hyperparameters for RL based controllers which may yield further performance gains.}{computationally constrained or partially observable environment models. However, under the assumptions that faults do not change the goal of a controller, the model is reliable, and sample sizes are unconstrained, model predictive control achieves faster recovery from faults.}


\begin{ack}
This work is supported by an Air Force SBIR Contract \# FA8501-14-P-0040 to Global Technology Connection (GTC), Inc., Atlanta, GA. Vanderbilt University has worked on this project under a subcontract from GTC, Inc.
\end{ack}
\bibliography{references}
\end{document}